\documentclass{article}

\usepackage{arxiv}

\usepackage[utf8]{inputenc} % allow utf-8 input
\usepackage[T1]{fontenc}    % use 8-bit T1 fonts
\usepackage[hidelinks]{hyperref}
\usepackage{graphicx}
\usepackage{caption}
\usepackage{subcaption}
\usepackage{url}            % simple URL typesetting
\usepackage{booktabs}       % professional-quality tables
\usepackage{amsfonts}       % blackboard math symbols
\usepackage{nicefrac}       % compact symbols for 1/2, etc.
\usepackage{microtype}      % microtypography
\usepackage{lipsum}		% Can be removed after putting your text content
\usepackage[round]{natbib}
\usepackage{doi}
\usepackage{amsthm}
\usepackage{amsmath}
\usepackage{mathabx}
\usepackage{diagbox}
\usepackage{stackengine}
\usepackage{algorithm}
\usepackage{algpseudocode}
 \usepackage{setspace}
\let\Algorithm\algorithm
\renewcommand\algorithm[1][]{\Algorithm[#1]\setstretch{1.2}}

\usepackage{array}
\usepackage{eqparbox}

\title{Optimizing Conversational Product Recommendation via Reinforcement Learning}

\date{} 					% Or removing it

\author{ 
\href{https://orcid.org/0000-0003-2002-984X}{\includegraphics[scale=0.06]{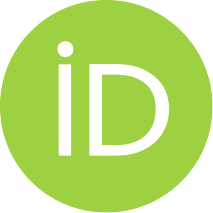}\hspace{1mm}Kang Liu} \\
Independent Researcher\\
\texttt{liukangk11@gmail.com} \\
}

\renewcommand{\undertitle}{}
\hypersetup{
pdftitle={A template for the arxiv style},
pdfsubject={q-bio.NC, q-bio.QM},
pdfauthor={David S.~Hippocampus, Elias D.~Striatum},
pdfkeywords={First keyword, Second keyword, More},
}

\begin{document}
\maketitle

\vspace{-20pt}

\begin{abstract}
We propose a reinforcement learning-based approach to optimize conversational strategies for product recommendation across diverse industries. As organizations increasingly adopt intelligent agents to support sales and service operations, the effectiveness of a conversation hinges not only on what is recommended but how and when recommendations are delivered. We explore a methodology where agentic systems learn optimal dialogue policies through feedback-driven reinforcement learning. By mining aggregate behavioral patterns and conversion outcomes, our approach enables agents to refine talk tracks that drive higher engagement and product uptake, while adhering to contextual and regulatory constraints. We outline the conceptual framework, highlight key innovations, and discuss the implications for scalable, personalized recommendation in enterprise environments.
\end{abstract}

% keywords can be removed
% \keywords{Reinforcement learning \and Simulation environment \and Cloud computing}

\section{Background}
Artificial intelligence (AI) has been transforming a wide range of industries, including robotics \citep{abiri2017tensile,liu2019vehicle}, manufacturing \citep{alavian2018alpha,alavian2019alpha,alavian2020alpha,alavian2022alpha,liu2021alpha,liu2021b,liu2025moment,liu2025general,eun2022production}, and healthcare \citep{jiang2017artificial,saraswat2022explainable}, by enabling automation, precision, and adaptive decision-making \citep{duan2019artificial,liu2024setcse}. This technological shift is now increasingly influencing consumer-facing domains—particularly marketing and customer service—where personalization and scalability are critical. Generative AI models, capable of producing fluent and context-aware text, are being leveraged to automate customer interactions, assist service agents, and tailor product recommendations in real time. As a result, there is a growing demand for AI systems that can not only generate responses, but also learn to optimize those responses to meet user needs and business objectives. Reinforcement learning offers a promising path forward by enabling agents to adapt and improve their conversational strategies based on real-world outcomes.

\section{Introduction}
The proliferation of conversational AI opens new possibilities for personalized product recommendation in sectors ranging from e-commerce and telecommunications to healthcare and education. While traditional recommender systems focus on matching user profiles to products or services, conversational settings introduce a temporal and strategic dimension: the path to recommendation matters. An agent must decide \textit{what to say, when to say it, and how to frame the offer}—all in the context of an evolving dialogue \citep{henderson2014second,liu2018dialogue,liu2020towards,lei2020estimation,adiwardana2020towards,jin2023amazon,li2024survey}.

Unlike static recommendation systems, conversational agents engage in real-time decision-making, where the success of a recommendation depends not only on the relevance of the product but also on the trajectory of interaction leading to the offer. This creates an opportunity to design agents that adapt their language, timing, and strategy based on user feedback across multiple turns.

Reinforcement learning (RL) \citep{Sutton1998} provides a natural foundation for modeling this sequential decision-making process. In this paradigm, the agent iteratively interacts with users (or user simulations), receives reward signals based on conversational outcomes, and learns policies that maximize long-term value—e.g., successful product conversions, user satisfaction, or adherence to operational guidelines \citep{young2013pomdp,li2016deep,zhao2016towards,dhingra2016towards,gao2023survey}.

Moreover, the ability to incorporate aggregate user response patterns and anonymized conversion data enables the design of privacy-conscious agents that generalize well across demographics and use cases. By integrating language models with RL, we open the door to learning robust dialogue strategies that evolve with customer needs, product offerings, and engagement goals.

We present a system-level approach to enabling such agents to optimize their strategies by learning from historical interactions and aggregate feedback, rather than explicit supervision or scripted logic.

\section{Problem Setting}
% The core challenge is to determine the optimal sequence of conversational acts (also called "talk tracks") that lead to successful product recommendations. The learning agent is modeled as a policy over states (dialogue contexts) and actions (utterance candidates). Feedback is received in the form of outcome-based rewards, such as conversion likelihood, engagement signals, and penalty for verbosity or non-compliance.

% Crucially, the reward is not always binary or deterministic. Instead, agents may receive aggregate feedback—e.g., most frequent customer responses and conversion rates observed historically in similar contexts. This allows for robust learning while preserving privacy and avoiding overfitting to specific cases.

The core challenge is to determine the optimal sequence of conversational acts (also called "talk tracks") that lead to successful product recommendations. In a conversational setting, each agent action not only affects the immediate response from the user but also shapes the context and tone for subsequent interactions. Therefore, the decision process must be treated as sequential and contextual, rather than as isolated prediction tasks.

We model this interaction as a Markov Decision Process (MDP), where:
\begin{itemize}
    \item \textbf{States} are dynamic dialogue contexts encoded using natural language embeddings.
    \item \textbf{Actions} are utterance candidates chosen from a pool of responses or generated by a language model.
    \item \textbf{Transitions} are governed by the user’s reply, which leads to the next dialogue state.
    \item \textbf{Rewards} reflect short-term or long-term signals such as engagement, clarity, helpfulness, or ultimate conversion.
\end{itemize}

Unlike conventional reinforcement learning environments, feedback in this domain is often sparse or noisy. Moreover, the final reward (e.g., whether a customer purchased the recommended product) may be delayed until the end of the conversation. To address this, we incorporate mechanisms to use aggregated historical outcomes, such as most frequent user responses and average conversion rates from prior sessions, rather than relying on exact ground-truth responses. This facilitates robust learning and supports privacy-aware modeling.

Additionally, dialogue environments may be subject to constraints and contextual signals, such as:
\begin{itemize}
\item Maximum number of dialogue turns before disengagement
\item Eligibility filters based on user type or product availability
\item Business rules and regulatory compliance considerations
\end{itemize}

These constraints and domain-specific characteristics must be accounted for in the design of both the learning agent and the surrounding infrastructure.

\section{Learning Framework}
% The agent operates in an episodic environment, where each episode corresponds to a simulated or logged customer conversation. At each turn:

% \begin{enumerate}
%     \item The agent observes the dialogue state (e.g., prior utterances, filled slots, estimated intent).
%     \item It selects a candidate action (e.g., a clarifying question, a direct offer).
%     \item It receives a feedback signal derived from aggregate customer behavior.
%     \item The agent updates its policy via standard RL techniques (e.g., Q-learning, policy gradients).
% \end{enumerate}

% To ensure safe and business-aligned learning, agents are guided by additional constraints, including:
% \begin{itemize}
%     \item Compliance filters that invalidate policy actions violating regulatory rules
%     \item Eligibility and context features from product metadata
%     \item Dialogue length or turn-budget limits
% \end{itemize}

% This setup mimics real-world constraints while preserving the agent’s capacity for strategy discovery.
The proposed learning framework treats each dialogue session as a multi-step decision process, where an agent must choose the most effective utterance at each turn to drive toward a final objective—such as product adoption, lead generation, or customer satisfaction. This episodic reinforcement learning setup allows the agent to incrementally improve its strategy by interacting with real users or simulated environments.

\subsection{State and Action Representation}
The state at each time step is derived from the full or partial conversation history, encoded using a language model (e.g., BERT or GPT) to capture semantic, syntactic, and contextual cues. The action space is a predefined set of possible agent utterances or templates, though it may also include generated responses in more advanced implementations.

\subsection{Environment and Reward}
The environment includes a response generator or simulator, which returns the next user utterance and a reward. The reward signal can be immediate (e.g., positive sentiment or continued engagement) or delayed (e.g., final product conversion or survey score). In our implementation, we allow the environment to return aggregated historical feedback (e.g., average conversion rates and most frequent responses) to simulate realistic outcomes while ensuring privacy.

\subsection{Policy Learning}
The agent learns a policy $\pi(a|s)$, which maps a given state (dialogue context) to a probability distribution over possible actions. We support multiple classes of RL algorithms for policy learning, including value-based (e.g., DQN), policy-gradient (e.g., PPO), and human-aligned (e.g., RLHF) approaches. These methods update parameters based on either sampled rewards or learned reward models.

\subsection{Safety and Compliance Constraints}
Before executing any action, it is passed through a compliance layer to ensure that it does not violate industry regulations, internal business rules, or ethical standards. If an action fails the compliance check, a fallback response is used or the action is masked during learning. This mechanism ensures that agents can safely explore without causing reputational or legal risks.

\subsection{Turn and Session Constraints}
The agent is also informed of session-level constraints such as the maximum number of remaining turns, which helps prioritize high-impact actions earlier in the conversation. These constraints help simulate realistic operational settings and make the learned policy more effective in production.

\subsection{Training Modes: Online and Offline}
\begin{itemize}
\item Offline Learning: The agent is trained on historical logs of customer-agent interactions. This data is anonymized and aggregated to ensure privacy.
\item Online Learning: The agent continues to refine its policy during live deployment, guided by real-time reward feedback and continual compliance checking.
\end{itemize}

This hybrid training setup enables scalable deployment and rapid adaptation to changing user behavior, product features, or market conditions.

\section{Algorithm}
We explore three core reinforcement learning algorithms for optimizing conversational recommendation strategies: Deep Q-Network (DQN) \citep{roderick2017implementing,kumar2020conservative}, Proximal Policy Optimization (PPO) \citep{schulman2017proximal}, and Reinforcement Learning with Human Feedback (RLHF) \citep{christiano2017deep,stiennon2020learning,bai2022training,ouyang2022training}. Each method offers distinct advantages depending on the feedback available, the complexity of the action space, and the desired level of policy control or alignment.

\subsection{Deep Q-Network}
DQN is a value-based method ideal for discrete action spaces, where the agent learns a Q-function that estimates the expected cumulative reward for each state-action pair. In our conversational setting, actions correspond to utterance templates or generated responses, and Q-values guide the agent toward the most rewarding sequences. The algorithm pseudo code is presented in Algorithm \ref{alg:dqn}.

\begin{algorithm}
\caption{DQN for optimal talking track}\label{alg:dqn}
\hspace*{\algorithmicindent} \texttt{Initialize Q-network $Q(s, a; \theta)$} \\
\hspace*{\algorithmicindent} \texttt{Initialize target network $Q_{\text{target}}(s, a; \bar{\theta})$} \\
\hspace*{\algorithmicindent} \texttt{Initialize replay buffer $D$}

\begin{algorithmic}[1]
\For{\texttt{episode in range(num\_episode)} }
    \State \texttt{Initialize conversation history $s_{\Theta}$}
    \For{\texttt{t in range(max\_turns)}} 
        \State \texttt{With probability $\epsilon$: choose random action $a_t$}
        \State \texttt{Else: $a_t \gets$ argmax\_a $Q(s_t, a; \theta)$}

        \State \texttt{Execute action $a_t \gets$ observe response, reward $r_t$, next state $s_{t+1}$}
        \State \texttt{Store $(s_t, a_t, r_t, s_{t+1})$ in $D$}

        \State \texttt{Sample minibatch from D:}
        \For{\texttt{each $(s_i, a_i, r_i, s_i')$}}
            \State $y_i \gets r_i + \gamma * \text{max\_{$a'$} } Q_{\text{target}}(s_i', a'; \bar{\theta})$
            \State $L \gets (Q(s_i, a_i; \theta) - y_i)^2$
        \EndFor
        \State \texttt{Update $\theta$ by minimizing loss $L$}
        \State \texttt{Periodically: $\bar{\theta} \gets \theta$ (update target network)}
        \State \texttt{Decay $\epsilon$}
    \EndFor
\EndFor
\end{algorithmic}
\end{algorithm}

\subsection{Proximal Policy Optimization}
PPO is a policy gradient method that strikes a balance between learning efficiency and stability. It directly learns a policy network that maps dialogue states to action probabilities, while also training a value network to estimate future rewards. PPO is especially useful when the action space is large or continuous, or when smooth policy updates are needed to avoid erratic behaviors. The algorithm pseudo code is presented in Algorithm \ref{alg:ppo}.

\begin{algorithm}
\caption{PPO for optimal talking track}\label{alg:ppo}
\hspace*{\algorithmicindent} \texttt{Initialize policy network $\pi(a|s; \theta)$} \\
\hspace*{\algorithmicindent} \texttt{Initialize value network $V(s; \phi)$}

\begin{algorithmic}[1]
\For{\texttt{iteration in range(num\_iterations)} }
    \State \texttt{Collect a batch of trajectories using $\pi$}
    \For{\texttt{t in range(max\_turns)}}
        \For{\texttt{t in range(max\_turns)}}
        \State \texttt{Sample action $a_t \sim \pi(a|s_t)$}
        \State \texttt{Execute $a_t \to$ observe reward $r_t$ and next state $s_{t+1}$}
        \State \texttt{Store $\bigl(s_t, a_t, r_t, \pi(a_t|s_t), V(s_t)\bigr)$}
        \EndFor
    \EndFor    
    \State \texttt{Compute advantages $A_t$ using GAE: $A_t \gets r_t + \gamma V(s_{t+1}) - V(s_t)$}
    \For{\texttt{K epochs}}
        \For{\texttt{For each batch of $(s_t, a_t, A_t, \text{old\_probs})$}}
            \State \texttt{Compute ratio $r_t(\theta) = \pi_{\theta}(a_t|s_t) / \text{old\_probs} $}
            \State \texttt{$L_{\text{clip}} \gets \min\bigl(r_t * A_t, \text{clip}(r_t, 1-\epsilon, 1+\epsilon)*A_t\bigr)$}
            \State \texttt{$L_{\text{value}} \gets \bigl(V(s_t; \phi) - R_t\bigr)^2$}
            \State \texttt{$L_{\text{entropy}} \gets -H\bigl(\pi(a|s_t)\bigr)$}
            \State \texttt{Update $\theta, \phi$ to maximize $L_{\text{clip}} + \beta * L_{entropy} - \lambda * L_{\text{value}}$}
        \EndFor
    \EndFor
\EndFor
\end{algorithmic}
\end{algorithm}

\subsection{Reinforcement Learning with Human Feedback}
RLHF augments standard reinforcement learning with a reward model trained on human preferences. It enables agents to align their responses more closely with human judgment, even when explicit reward signals are sparse or noisy. In the context of conversational recommendations, RLHF is particularly valuable when fine-tuning language models to balance persuasiveness, helpfulness, and safety. The algorithm pseudo code is presented in Algorithm \ref{alg:rlhf}.

\begin{algorithm}
\caption{RLHF for optimal talking track}\label{alg:rlhf}

\texttt{\# Step 1: Supervised Fine-Tuning} \\
\texttt{Train base policy $\pi_{\theta}(a|s)$ on human-annotated dialogues} \\

\texttt{\# Step 2: Reward Model Training}\\
\texttt{Collect (prompt, response A, response B, human preference) tuples}\\
\texttt{Train reward model $R(s, a)$ to assign higher scores to preferred responses}\\

\texttt{\# Step 3: Policy Fine-Tuning with PPO}\\
\texttt{Initialize policy $\pi(a|s; \theta) \gets \pi_{\Theta}$}
\begin{algorithmic}[1]
\For{\texttt{iteration in range(num\_iterations)} }
    \For{\texttt{each prompt in $s$}}
        \State \texttt{Generate response $a \sim \pi$}
        \State \texttt{Compute reward $r \gets R(s, a)$}
        \State \texttt{Store $\bigl(s, a, r, \pi(a|s)\bigr)$ in buffer}
    \EndFor
    \State \texttt{Compute advantages $A_t$ from rewards (e.g., using baseline or value function)}
    \For{\texttt{each batch}}
        \State \text{Compute $r_t = \pi(a_t|s_t) / text{old_probs}$}
        \State \texttt{$L_{\text{clip}} \gets \min\bigl( r_t*A_t, clip(r_t, 1 - \epsilon, 1+ \epsilon)*A_t \bigr)$}
        \State \texttt{Update $\theta$ to maximize $L_{\text{clip}}$}
    \EndFor    
\EndFor
\end{algorithmic}
\end{algorithm}

\section{Key Contributions}
Our work introduces a reinforcement learning-based methodology that not only automates product recommendation but also discovers effective communication strategies over multiple dialogue turns. By leveraging recent advances in generative modeling, sequential decision-making, and privacy-preserving training techniques, our approach creates a unified framework for deploying scalable, intelligent agents across industries. The following contributions highlight our technical and conceptual innovations:

\begin{itemize}
    \item \textbf{Talk Track Optimization}: We frame product recommendation as a sequential decision problem and apply RL to learn dynamic strategies.

    \item \textbf{Aggregate Reward Modeling}: Instead of relying on individual labels, we use statistical aggregates (e.g., most frequent user replies, conversion rate averages) to train policies in a privacy-conscious manner.

    \item \textbf{Constraint-Aware Exploration}: Policies are learned within a bounded space defined by regulatory, business, and interaction constraints, ensuring safe deployment readiness.

    \item \textbf{Scalability}: Our framework supports diverse agents learning from varied product lines, customer segments, and interaction histories.
\end{itemize}

\section{Implications and Future Work}
This approach offers a foundation for goal-directed, adaptive recommendation systems in any domain that involves sequential, interactive communication. By learning not just what to recommend but how to recommend it, agents can become more persuasive, trustworthy, and aligned with business goals. The use of reinforcement learning enables these systems to adapt to evolving user preferences, changing product catalogs, and variable conversation constraints.

Organizations can use this framework to:
\begin{itemize}
\item Improve conversion rates via personalized, optimized talk tracks
\item Reduce the need for human agent interventions through more effective AI interactions
\item Achieve regulatory and ethical alignment by integrating compliance checks directly into the learning loop
\item Leverage historical interaction logs in a privacy-aware fashion to train and improve models without compromising data security
\item Facilitate A/B testing and longitudinal improvement by comparing different policy versions in production environments
\end{itemize}

Future work includes expanding the framework along several directions:
\begin{itemize}
\item Incorporating simulation environments to enable safe pre-deployment training and evaluation
\item Applying human-in-the-loop reinforcement learning to capture nuanced user preferences beyond conversion metrics
\item Using meta-learning or curriculum learning strategies to accelerate agent adaptation in new domains or product categories
\item Designing richer reward functions that balance user satisfaction, engagement, and operational efficiency
\item Extending the framework to multi-agent settings, where cooperative or competitive agent behavior may arise (e.g., multi-brand product catalogs)
\end{itemize}

These directions can further enhance the generalizability, robustness, and effectiveness of agentic recommendation systems across industries.

\section{Conclusion}
Reinforcement learning offers a powerful lens through which to design intelligent agents that go beyond matching products to users—they learn how to \textit{communicate effectively} to achieve that goal. By leveraging historical dialogue data and aggregated feedback, we can build agents that discover optimal talk tracks for personalized recommendation, aligning user outcomes with organizational goals in a responsible and scalable manner.

The integration of RL with natural language processing and compliance-aware design enables a new class of conversational agents that adapt over time, respond to real user signals, and maintain safety in high-stakes environments. These systems are not only technically viable but also practically impactful, especially as industries seek scalable and intelligent approaches to customer engagement.

We anticipate this work serving as a foundation for next-generation recommendation and servicing systems that can personalize at scale, learn continually from interactions, and uphold trust and transparency in human-AI communication.

\bibliographystyle{unsrtnat}
\bibliography{references}  %%% Uncomment this line and comment out the ``thebibliography'' section below to use the external .bib file (using bibtex) .

\begin{thebibliography}{35}
\providecommand{\natexlab}[1]{#1}
\providecommand{\url}[1]{\texttt{#1}}
\expandafter\ifx\csname urlstyle\endcsname\relax
  \providecommand{\doi}[1]{doi: #1}\else
  \providecommand{\doi}{doi: \begingroup \urlstyle{rm}\Url}\fi

\bibitem[Abiri et~al.(2017)Abiri, Paydar, Tao, LaRocca, Liu, Genovese, Candler, Grundfest, and Dutson]{abiri2017tensile}
Ahmad Abiri, Omeed Paydar, Anna Tao, Megan LaRocca, Kang Liu, Bradley Genovese, Robert Candler, Warren~S Grundfest, and Erik~P Dutson.
\newblock Tensile strength and failure load of sutures for robotic surgery.
\newblock \emph{Surgical endoscopy}, 31:\penalty0 3258--3270, 2017.

\bibitem[Liu et~al.(2019)Liu, Li, Kolmanovsky, and Girard]{liu2019vehicle}
Kang Liu, Nan Li, Ilya Kolmanovsky, and Anouck Girard.
\newblock A vehicle routing problem with dynamic demands and restricted failures solved using stochastic predictive control.
\newblock In \emph{2019 American Control Conference (ACC)}, pages 1885--1890. IEEE, 2019.

\bibitem[Alavian et~al.(2018)Alavian, Eun, Liu, Meerkov, and Zhang]{alavian2018alpha}
Pooya Alavian, Yongsoon Eun, Kang Liu, Semyon~M Meerkov, and Liang Zhang.
\newblock The ($\alpha$, $\beta$)-precise estimates of mtbf and mttr: Definitions, calculations, and effect on machine efficiency and throughput evaluation in serial production lines.
\newblock \emph{URL: http://web. eecs. umich. edu/\~{} smm/publications/mtbf\_mttr\_estimates. pdf}, 2018.

\bibitem[Alavian et~al.(2019)Alavian, Eun, Liu, Meerkov, and Zhang]{alavian2019alpha}
Pooya Alavian, Yongsoon Eun, Kang Liu, Semyon~M Meerkov, and Liang Zhang.
\newblock The ($\alpha$, $\beta$)-precise estimates of mtbf and mttr: Definitions, calculations, and induced effect on machine efficiency evaluation.
\newblock \emph{IFAC-PapersOnLine}, 52\penalty0 (13):\penalty0 1004--1009, 2019.

\bibitem[Alavian et~al.(2020)Alavian, Eun, Liu, Meerkov, and Zhang]{alavian2020alpha}
Pooya Alavian, Yongsoon Eun, Kang Liu, Semyon~M Meerkov, and Liang Zhang.
\newblock The ($\alpha$, $\beta$)-precise estimates of mtbf and mttr: Definition, calculation, and observation time.
\newblock \emph{IEEE Transactions on Automation Science and Engineering}, 18\penalty0 (3):\penalty0 1469--1477, 2020.

\bibitem[Alavian et~al.(2022)Alavian, Eun, Liu, Meerkov, and Zhang]{alavian2022alpha}
Pooya Alavian, Yongsoon Eun, Kang Liu, Semyon~M Meerkov, and Liang Zhang.
\newblock The ($\alpha$ x, $\beta$ x)-precise estimates of production systems performance metrics.
\newblock \emph{International Journal of Production Research}, 60\penalty0 (7):\penalty0 2230--2253, 2022.

\bibitem[Liu(2021{\natexlab{a}})]{liu2021alpha}
K~Liu.
\newblock \emph{The ($\alpha$, $\beta$)-precision theory for production system monitoring and improvement}.
\newblock PhD thesis, Ph. D. thesis, The University of Michigan, 2021{\natexlab{a}}.

\bibitem[Liu(2021{\natexlab{b}})]{liu2021b}
Kang Liu.
\newblock \emph{The (a, b)-Precision Theory for Production System Monitoring and Improvement}.
\newblock PhD thesis, Ph. D. thesis, The University of Michigan, 2021{\natexlab{b}}.

\bibitem[Liu(2025{\natexlab{a}})]{liu2025moment}
Kang Liu.
\newblock Moment monotonicity of weibull, gamma and log-normal distributions.
\newblock \emph{arXiv preprint arXiv:2502.11366}, 2025{\natexlab{a}}.

\bibitem[Liu(2025{\natexlab{b}})]{liu2025general}
Kang Liu.
\newblock General form moment-based estimator of weibull, gamma, and log-normal distributions.
\newblock \emph{arXiv preprint arXiv:2505.01911}, 2025{\natexlab{b}}.

\bibitem[Eun et~al.(2022)Eun, Liu, and Meerkov]{eun2022production}
Yongsoon Eun, Kang Liu, and Semyon~M Meerkov.
\newblock Production systems with cycle overrun: modelling, analysis, improvability and bottlenecks.
\newblock \emph{International Journal of Production Research}, 60\penalty0 (2):\penalty0 534--548, 2022.

\bibitem[Jiang et~al.(2017)Jiang, Jiang, Zhi, Dong, Li, Ma, Wang, Dong, Shen, and Wang]{jiang2017artificial}
Fei Jiang, Yong Jiang, Hui Zhi, Yi~Dong, Hao Li, Sufeng Ma, Yilong Wang, Qiang Dong, Haipeng Shen, and Yongjun Wang.
\newblock Artificial intelligence in healthcare: past, present and future.
\newblock \emph{Stroke and vascular neurology}, 2\penalty0 (4), 2017.

\bibitem[Saraswat et~al.(2022)Saraswat, Bhattacharya, Verma, Prasad, Tanwar, Sharma, Bokoro, and Sharma]{saraswat2022explainable}
Deepti Saraswat, Pronaya Bhattacharya, Ashwin Verma, Vivek~Kumar Prasad, Sudeep Tanwar, Gulshan Sharma, Pitshou~N Bokoro, and Ravi Sharma.
\newblock Explainable ai for healthcare 5.0: opportunities and challenges.
\newblock \emph{IEEe Access}, 10:\penalty0 84486--84517, 2022.

\bibitem[Duan et~al.(2019)Duan, Edwards, and Dwivedi]{duan2019artificial}
Yanqing Duan, John~S Edwards, and Yogesh~K Dwivedi.
\newblock Artificial intelligence for decision making in the era of big data--evolution, challenges and research agenda.
\newblock \emph{International journal of information management}, 48:\penalty0 63--71, 2019.

\bibitem[Liu(2024)]{liu2024setcse}
Kang Liu.
\newblock Setcse: Set operations using contrastive learning of sentence embeddings.
\newblock \emph{arXiv preprint arXiv:2404.17606}, 2024.

\bibitem[Henderson et~al.(2014)Henderson, Thomson, and Williams]{henderson2014second}
Matthew Henderson, Blaise Thomson, and Jason~D Williams.
\newblock The second dialog state tracking challenge.
\newblock In \emph{Proceedings of the 15th annual meeting of the special interest group on discourse and dialogue (SIGDIAL)}, pages 263--272, 2014.

\bibitem[Liu et~al.(2018)Liu, Tur, Hakkani-Tur, Shah, and Heck]{liu2018dialogue}
Bing Liu, Gokhan Tur, Dilek Hakkani-Tur, Pararth Shah, and Larry Heck.
\newblock Dialogue learning with human teaching and feedback in end-to-end trainable task-oriented dialogue systems.
\newblock \emph{arXiv preprint arXiv:1804.06512}, 2018.

\bibitem[Liu et~al.(2020)Liu, Wang, Niu, Wu, Che, and Liu]{liu2020towards}
Zeming Liu, Haifeng Wang, Zheng-Yu Niu, Hua Wu, Wanxiang Che, and Ting Liu.
\newblock Towards conversational recommendation over multi-type dialogs.
\newblock \emph{arXiv preprint arXiv:2005.03954}, 2020.

\bibitem[Lei et~al.(2020)Lei, He, Miao, Wu, Hong, Kan, and Chua]{lei2020estimation}
Wenqiang Lei, Xiangnan He, Yisong Miao, Qingyun Wu, Richang Hong, Min-Yen Kan, and Tat-Seng Chua.
\newblock Estimation-action-reflection: Towards deep interaction between conversational and recommender systems.
\newblock In \emph{Proceedings of the 13th international conference on web search and data mining}, pages 304--312, 2020.

\bibitem[Adiwardana et~al.(2020)Adiwardana, Luong, So, Hall, Fiedel, Thoppilan, Yang, Kulshreshtha, Nemade, Lu, et~al.]{adiwardana2020towards}
Daniel Adiwardana, Minh-Thang Luong, David~R So, Jamie Hall, Noah Fiedel, Romal Thoppilan, Zi~Yang, Apoorv Kulshreshtha, Gaurav Nemade, Yifeng Lu, et~al.
\newblock Towards a human-like open-domain chatbot.
\newblock \emph{arXiv preprint arXiv:2001.09977}, 2020.

\bibitem[Jin et~al.(2023)Jin, Mao, Li, Jiang, Luo, Wen, Han, Lu, Wang, Li, et~al.]{jin2023amazon}
Wei Jin, Haitao Mao, Zheng Li, Haoming Jiang, Chen Luo, Hongzhi Wen, Haoyu Han, Hanqing Lu, Zhengyang Wang, Ruirui Li, et~al.
\newblock Amazon-m2: A multilingual multi-locale shopping session dataset for recommendation and text generation.
\newblock \emph{Advances in Neural Information Processing Systems}, 36:\penalty0 8006--8026, 2023.

\bibitem[Li et~al.(2024)Li, Lin, Wang, Feng, Pang, Li, Nie, He, and Chua]{li2024survey}
Yongqi Li, Xinyu Lin, Wenjie Wang, Fuli Feng, Liang Pang, Wenjie Li, Liqiang Nie, Xiangnan He, and Tat-Seng Chua.
\newblock A survey of generative search and recommendation in the era of large language models.
\newblock \emph{arXiv preprint arXiv:2404.16924}, 2024.

\bibitem[Sutton and Barto(2018)]{Sutton1998}
Richard~S. Sutton and Andrew~G. Barto.
\newblock \emph{Reinforcement Learning: An Introduction}.
\newblock The MIT Press, second edition, 2018.
\newblock URL \url{http://incompleteideas.net/book/the-book-2nd.html}.

\bibitem[Young et~al.(2013)Young, Ga{\v{s}}i{\'c}, Thomson, and Williams]{young2013pomdp}
Steve Young, Milica Ga{\v{s}}i{\'c}, Blaise Thomson, and Jason~D Williams.
\newblock Pomdp-based statistical spoken dialog systems: A review.
\newblock \emph{Proceedings of the IEEE}, 101\penalty0 (5):\penalty0 1160--1179, 2013.

\bibitem[Li et~al.(2016)Li, Monroe, Ritter, Galley, Gao, and Jurafsky]{li2016deep}
Jiwei Li, Will Monroe, Alan Ritter, Michel Galley, Jianfeng Gao, and Dan Jurafsky.
\newblock Deep reinforcement learning for dialogue generation.
\newblock \emph{arXiv preprint arXiv:1606.01541}, 2016.

\bibitem[Zhao and Eskenazi(2016)]{zhao2016towards}
Tiancheng Zhao and Maxine Eskenazi.
\newblock Towards end-to-end learning for dialog state tracking and management using deep reinforcement learning.
\newblock \emph{arXiv preprint arXiv:1606.02560}, 2016.

\bibitem[Dhingra et~al.(2016)Dhingra, Li, Li, Gao, Chen, Ahmed, and Deng]{dhingra2016towards}
Bhuwan Dhingra, Lihong Li, Xiujun Li, Jianfeng Gao, Yun-Nung Chen, Faisal Ahmed, and Li~Deng.
\newblock Towards end-to-end reinforcement learning of dialogue agents for information access.
\newblock \emph{arXiv preprint arXiv:1609.00777}, 2016.

\bibitem[Gao et~al.(2023)Gao, Zheng, Li, Li, Qin, Piao, Quan, Chang, Jin, He, et~al.]{gao2023survey}
Chen Gao, Yu~Zheng, Nian Li, Yinfeng Li, Yingrong Qin, Jinghua Piao, Yuhan Quan, Jianxin Chang, Depeng Jin, Xiangnan He, et~al.
\newblock A survey of graph neural networks for recommender systems: Challenges, methods, and directions.
\newblock \emph{ACM Transactions on Recommender Systems}, 1\penalty0 (1):\penalty0 1--51, 2023.

\bibitem[Roderick et~al.(2017)Roderick, MacGlashan, and Tellex]{roderick2017implementing}
Melrose Roderick, James MacGlashan, and Stefanie Tellex.
\newblock Implementing the deep q-network.
\newblock \emph{arXiv preprint arXiv:1711.07478}, 2017.

\bibitem[Kumar et~al.(2020)Kumar, Zhou, Tucker, and Levine]{kumar2020conservative}
Aviral Kumar, Aurick Zhou, George Tucker, and Sergey Levine.
\newblock Conservative q-learning for offline reinforcement learning.
\newblock \emph{Advances in neural information processing systems}, 33:\penalty0 1179--1191, 2020.

\bibitem[Schulman et~al.(2017)Schulman, Wolski, Dhariwal, Radford, and Klimov]{schulman2017proximal}
John Schulman, Filip Wolski, Prafulla Dhariwal, Alec Radford, and Oleg Klimov.
\newblock Proximal policy optimization algorithms.
\newblock \emph{arXiv preprint arXiv:1707.06347}, 2017.

\bibitem[Christiano et~al.(2017)Christiano, Leike, Brown, Martic, Legg, and Amodei]{christiano2017deep}
Paul~F Christiano, Jan Leike, Tom Brown, Miljan Martic, Shane Legg, and Dario Amodei.
\newblock Deep reinforcement learning from human preferences.
\newblock \emph{Advances in neural information processing systems}, 30, 2017.

\bibitem[Stiennon et~al.(2020)Stiennon, Ouyang, Wu, Ziegler, Lowe, Voss, Radford, Amodei, and Christiano]{stiennon2020learning}
Nisan Stiennon, Long Ouyang, Jeffrey Wu, Daniel Ziegler, Ryan Lowe, Chelsea Voss, Alec Radford, Dario Amodei, and Paul~F Christiano.
\newblock Learning to summarize with human feedback.
\newblock \emph{Advances in neural information processing systems}, 33:\penalty0 3008--3021, 2020.

\bibitem[Bai et~al.(2022)Bai, Jones, Ndousse, Askell, Chen, DasSarma, Drain, Fort, Ganguli, Henighan, et~al.]{bai2022training}
Yuntao Bai, Andy Jones, Kamal Ndousse, Amanda Askell, Anna Chen, Nova DasSarma, Dawn Drain, Stanislav Fort, Deep Ganguli, Tom Henighan, et~al.
\newblock Training a helpful and harmless assistant with reinforcement learning from human feedback.
\newblock \emph{arXiv preprint arXiv:2204.05862}, 2022.

\bibitem[Ouyang et~al.(2022)Ouyang, Wu, Jiang, Almeida, Wainwright, Mishkin, Zhang, Agarwal, Slama, Ray, et~al.]{ouyang2022training}
Long Ouyang, Jeffrey Wu, Xu~Jiang, Diogo Almeida, Carroll Wainwright, Pamela Mishkin, Chong Zhang, Sandhini Agarwal, Katarina Slama, Alex Ray, et~al.
\newblock Training language models to follow instructions with human feedback.
\newblock \emph{Advances in neural information processing systems}, 35:\penalty0 27730--27744, 2022.

\end{thebibliography}

\end{document}